%% file: paper.tex
\documentclass[twocolumn,showpacs,aps,floatfix,prd,nofootinbib]{revtex4-1}

\usepackage{graphicx}
\usepackage{dcolumn}
\usepackage{array, longtable}
\usepackage{epsfig}
\usepackage{rotating} 
\usepackage{amsmath}

\newcommand{\BaBarYear}       {11}
\newcommand{\BaBarNumber}     {016}
\newcommand{\SLACPubNumber} {14857}
\newcommand{\BaBarType}      {PUB}  

\long\def\inst#1{\par\nobreak\kern 4pt\nobreak
    {\it #1}\par\vskip 10pt plus 3pt minus 3pt}

\input{babarsym}

\begin{document}

\begin{flushleft}
\babar-\BaBarType-\BaBarYear/\BaBarNumber \\
SLAC-PUB-\SLACPubNumber \\
PRD 85, 112009 (2012)\\
\vspace{0.14cm}
\end{flushleft}

\title{
\large \bf \boldmath
Initial-State Radiation Measurement of the $\ep\en\to\pip\pim\pip\pim$ Cross Section
}

\input authors_jul2011.tex

\begin{abstract}
We study the process $e^+e^-\to\pip\pim\pip\pim\g$, with a photon emitted from the initial-state electron or positron, using 454.3\invfb of data collected with the \babar\ detector at SLAC, corresponding to approximately $260\,000$ signal events. We use these data to extract the non-radiative $\sigma(e^+e^-\to\pip\pim\pip\pim)$ cross section in the energy range from 0.6 to 4.5 \gev. The total uncertainty of the cross section measurement in the peak region is less than 3\%, higher in precision than the corresponding results obtained from energy scan data.
\end{abstract}


\maketitle

\setcounter{footnote}{0}

\section{Introduction}
\label{sec:Introduction}

While the electroweak contribution to the anomalous magnetic moment of the muon $a_{\mu}=\frac{1}{2}(g_\mu-2)$ can be calculated with sub-ppm precision, the hadronic contribution $a^{\mathrm{had}}_\mu$ cannot be evaluated by means of perturbative techniques within Quantum Chromodynamics (QCD). However, it is possible to relate $a^{\mathrm{had}}_\mu$ via a dispersion relation to hadronic cross section $\sigma(e^+e^-\to $ hadrons$)$ data~\cite{cabibo}. Thus, hadronic cross section measurements are important for the Standard Model prediction of $a_{\mu}$, which differs by more than three standard deviations from a recent direct BNL measurement~\cite{BNL}.

The study of initial-state radiation (ISR) events at an $\ep\en$ collider with a fixed center-of-mass (CM) energy $E_{CM}$ allows high-precision measurements of the cross section of exclusive hadronic channels for energies below the nominal $E_{CM}$ and is complementary to studies based on an energy scan. Use of the ISR technique is discussed in Refs.~\cite{arbus, kuehn, ivanch,eidelman}. Previously, we used this technique to investigate low-multiplicity hadronic processes at effective CM~energies below 5\gev~\cite{3pi, 4piold, 6pi, pp, lam, 5pi, 2k2pi_new, 2kpi, 2pi}. 

The ISR cross section $\sigma_{f,\g}$ for a specific final state $f$ depends on the non-radiative cross section $\sigma_{f}$ and is given by~\cite{ivanch}: 

\begin{equation}
\frac{d\sigma_{f,\g} (s, M_{had})}{dM_{had}} = \frac{2M_{had}}{s}\cdot W(s,x,C^{\ast})\cdot \sigma_{f}(M_{had}),
\label{radxs}
\end{equation}
where $x=2\cdot E^{\ast}_{\g}/\sqrt{s}$, $\sqrt{s}$ is the nominal CM~energy, $E^{\ast}_{\g}$ is the energy of the ISR photon,\footnote{$^\ast$ refers to the nominal CM frame.} and $M_{had}=\sqrt{s(1-x)}$. The radiator function $W(s,x,C^{\ast})$, which describes the probability for the emission of an ISR photon in the polar angle range $|\cos\theta^{\ast}_{\g}| < C^{\ast}$, is determined to next-to-leading order (NLO) ISR with the PHOKHARA software package~\cite{phok}. $\theta^{\ast}$ is the angle of the emitted $\gamma$ with respect to the direction of the incoming $\en$. Effects of final-state radiation (FSR), which are neglected in eq.~(\ref{radxs}), are studied and corrected for as explained in section~\ref{sec:xs}.

Recently, cross section measurements  with approximately $1\%$ precision have been presented for the $\pip\pim$ final state~\cite{2pi,kloe2pi,kloe2pi2}. These measurements have led to a significant reduction in the uncertainty of $a^{\mathrm{had}}_\mu$~\cite{davg-2}. As a consequence, a large relative contribution to the uncertainty of $a^{\mathrm{had}}_\mu$ now arises from the energy range $1\gev<E<2\gev$~\cite{jn}. In this energy range, the hadronic cross section is dominated by the exclusive $\ep\en\to\pip\pim\pip\pim$ and $\pip\pim\piz\piz$ channels. 

This paper reports results from an ISR analysis of the $\ep\en\to\pip\pim\pip\pim\gamma$ process. The four-pion mass $M_{4\pi}$ serves as the effective hadronic $E_{CM}$ value. It is an update of our earlier study~\cite{4piold}, based on a data sample that is five times larger. We perform a more detailed study of systematic effects and thereby obtain significant improvements in both the statistical and systematic uncertainties.

The outline of this paper is as follows. Section~\ref{sec:babar} describes the \babar\ detector and the data set used in this analysis. The primary event selection is presented in section~\ref{sec:evsel}. After discussing the background suppression (section~\ref{sec:bkg}) and the acceptance and efficiency studies (section~\ref{sec:acc}), the extraction of the non-radiative cross section is described in section~\ref{sec:xs}. A qualitative analysis of the intermediate subsystems and a quantitative measurement of the $\jpsi$ and $\psitwos$ branching fractions follows in section~\ref{sec:inv}. A summary is given in section~\ref{sec:Summary}.

\section{\boldmath The \babar\ detector and dataset}
\label{sec:babar}
The data used in this analysis were collected with the \babar\ detector at the \pep2\ asymmetric energy \epem\ storage rings at the SLAC National Accelerator Laboratory. A total integrated luminosity of 454.3 \invfb is used, comprised of 413.1\invfb collected at the \FourS resonance peak, and 41.2\invfb collected 40\mev below the peak.

The \babar\ detector is described in detail elsewhere~\cite{babar}. The reconstruction of charged-particle tracks is performed with the tracking system, which is comprised of a five-layer silicon vertex tracker (SVT) and a 40-layer drift chamber (DCH), both in a 1.5 T axial magnetic field. Separation of electrons, protons, charged pions, and charged kaons is achieved using Cherenkov angles measured with the detector of internally reflected Cherenkov light (DIRC) in combination with specific ionization \dedx measurements from the SVT and DCH. The CsI(Tl) electromagnetic calorimeter (EMC) measures the energy of photons and electrons. Muon identification is provided by the instrumented flux return.

A simulation package developed for radiative processes, AFKQED, is used to determine detector acceptance and reconstruction efficiencies. Hadronic final states, including $\pipi\pipi\g$, are simulated based on an approach of Czy\.{z} and K\"{u}hn~\cite{kuehn2}. The underlying model assumes dominance of the $a_{1}(1260)\pi$ final state as was reported in Refs.~\cite{4pi_cmd, Bondar}. Due to the dominant decay $a_{1}(1260)\to\rho^0\pi$, each event contains one pair of pions from $\rho^0$ decay. The simulation of multiple soft-photon emission from the initial state is performed via a structure function technique~\cite{kuraev,strfun}. Extra radiation from the final state particles is simulated by the PHOTOS package~\cite{PHOTOS}. The accuracy of the radiative corrections is about 1$\%$. A new version of PHOKHARA~\cite{phok}, which incorporates the results of recent studies~\cite{4piold} on intermediate resonances, is used to investigate the influence of these intermediate resonances on the acceptance.

The simulated events are subjected to simulation of the detector~\cite{geant} and the same analysis procedures as the data. Variations in detector and background conditions over the course of the experiment are modeled.

A large number of potential background processes are simulated, including the ISR processes $\Kp\Km\pip\pim\g$ and $\KS\Kpm\pimp\g$. Other ISR background channels are also examined. Either the remaining number of events after the event selection is negligible in comparison to other uncertainties, or dedicated methods for background rejection are implemented as described in section~\ref{sec:bkg}. Non-ISR backgrounds resulting from $\ep\en\to\qqbar$  (\q=\u,\d,\s,\c) continuum events are modeled using the JETSET generator~\cite{jetset}, while those from $\ep\en\to\taup\taum$ are modeled with KORALB~\cite{koralb}. The cross sections of the ISR channels and the branching fractions of the inclusive processes leading to final states similar to our signal are known with about 10$\%$ accuracy or better, which is sufficient for the purpose of this measurement. The contribution from $\FourS\to\pipi\pipi\g$ events is negligible.

\section{Primary event selection and kinematic fit}
\label{sec:evsel}

Charged tracks are selected by requiring that they originate from the collision region (transverse distance of closest approach to the nominal interaction point $d_{T}<1.5\cm$, and in the beam direction $\d_{Z}< 2.5\cm$), and that they have a polar angle $\theta_{ch}$ in the well-understood acceptance region of the detector ($0.5\rad<\theta_{ch}<2.4\rad$). The coordinate system has the z-axis in the direction of the incoming $\en$ beam. Tracks with transverse momenta less than 100\mevc or that are consistent with being an electron are rejected. Photon candidates are required to have a minimum energy $E_{\g, CM}>$ 50\mev. The ISR photon candidate is restricted to the polar angle range inside the well-understood acceptance region of the EMC ($0.35\rad<\theta_{\gamma}<2.4\rad$) and a minimum energy of $E_{\mathrm{ISR}}>3\gev$ is required.

Radiative Bhabha events are suppressed by requiring the two most energetic tracks of the event not to be identified as electrons. 
The minimum angle between a charged track and the ISR photon $\Delta\psi$ is required to satisfy $\Delta\psi>1.0\rad$, in order to select the back-to-back topology between the charged tracks and photon typical for ISR events. 

A kinematic fit procedure with four constraints (4C) is applied to events with four tracks satisfying these criteria. The constrained fit uses the measured momenta and directions of the charged particles and the ISR photon and the corresponding error matrix to solve the energy-momentum equation. To provide accurate photon parameters to the kinematic fit, a precise alignment and an energy calibration of the EMC are performed using a $\mup\mun\g$ sample. This improves the data-simulation agreement in the goodness-of-fit distributions (see below) of the kinematic fit. The $\mup\mun\g$ sample is also used to identify and measure differences between the data and the simulation (MC) in the photon detection efficiency.

The kinematic fit is performed assuming the $\pip\pim\pip\pim\gamma$ signal hypothesis. If two tracks are identified as kaons the fit is also performed under the $\Kp\Km\pip\pim\gamma$ hypothesis. The fitting routine returns a goodness-of-fit quantity $\chisq_{4\pi}$ and $\chisq_{2K2\pi}$, respectively, which we use to select signal events and to suppress $\Kp\Km\pip\pim\gamma$ background.

\begin{figure}[h]
\begin{center}
\includegraphics[clip, width=0.5\textwidth]{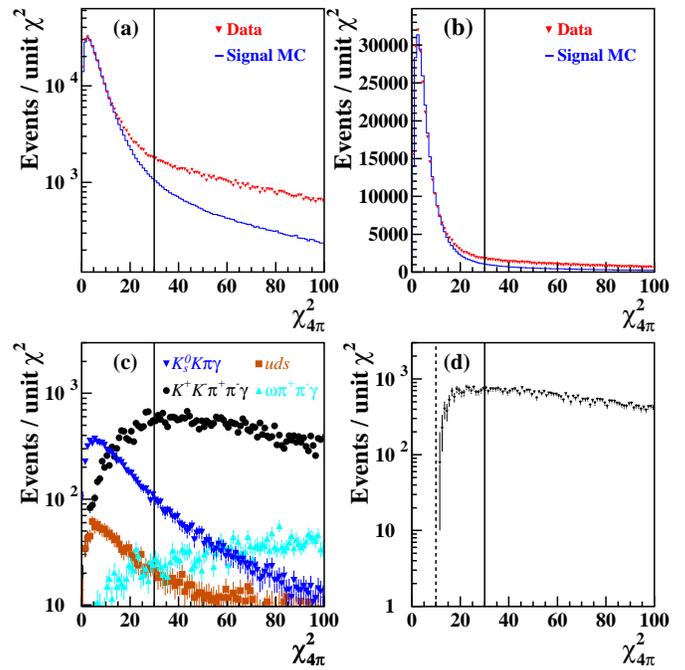}
\end{center}
\caption{(a,b) The $\chisq_{4\pi}$ distribution for data without background subtraction (triangles) and signal MC (histogram) on a logarithmic and linear scale. (c) MC distributions of the principal backgrounds. (d) The difference between the measured and the simulated distributions from part (a,b). The dashed and solid vertical lines indicate the boundaries with the $0<\chisq_{4\pi}<10$ region used for normalization and the $\chisq_{4\pi}<30$ requirement, respectively.}
\label{4pi_babar}
\end{figure}

Figures~\ref{4pi_babar}(a) and (b) show the measured $\chisq_{4\pi}$ distribution before background subtraction in comparison to the distribution from the simulated signal sample. The results are shown on logarithmic and linear scales.
The distributions are normalized to the number of events with $\chisq_{4\pi}<10$.
Figure~\ref{4pi_babar}(c) shows the $\chisq_{4\pi}$ distribution for the main backgrounds. 
The difference between measured and simulated samples is shown in Fig.~\ref{4pi_babar}(d) for $\chisq_{4\pi}>10$.  For large values of $\chisq_{4\pi}$, the difference is approximately flat, and is consistent with the sum of backgrounds shown in Fig.~\ref{4pi_babar}(c). At small $\chisq_{4\pi}$ values, resolution effects cause the difference to decrease. This is studied with a clean sample of four pions and one photon where the resolution effect is visible for $\chisq_{4\pi} < 20$. To avoid a bias due to the resolution difference between the data and simulation, the requirement $\chisq_{4\pi} < 30$ is used in the event analysis. In addition, background channels are suppressed using dedicated vetoes and particle identification (PID) selectors, as discussed in section~\ref{sec:bkg}.        

\section{Background Rejection}
\label{sec:bkg}
\begin{figure}[h]
\begin{center}
\includegraphics[clip, width=0.5\textwidth]{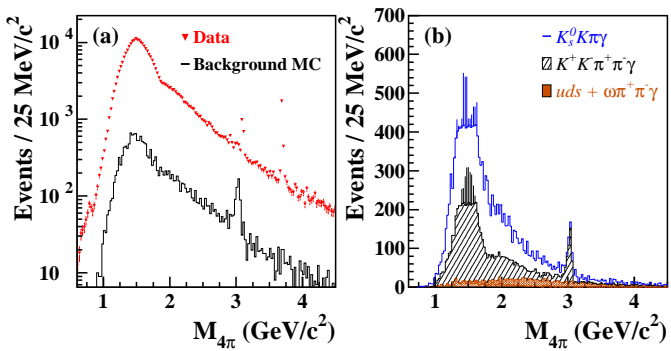}
\end{center}
\caption{(a) The $\pipi\pipi$ invariant mass distribution of $\pip\pim\pip\pim\gamma$ ISR events (triangles) for the main selection and the sum of simulated $\Kp\Km\pip\pim\gamma$, $\KS\Kpm\pimp\gamma$, $\omega\pip\pim\gamma$ and non-radiative $\u\d\s$-continuum background (histogram). The large $\pip\pim\gamma\gamma$ and $\pip\pim\piz\gamma$ background contribution in the threshold region ($M_{4\pi}<1.1\gevcc$) is not included in the simulation. (b) The individual simulated background channel contributions from top to bottom as indicated in the legend.}
\label{m4pi_babar}
\end{figure}

Due to large variations in the signal-to-background ratio across different 4$\pi$ mass $M_{4\pi}$ regions, different strategies are implemented to eliminate background, depending on the $M_{4\pi}$ region.
Figure~\ref{m4pi_babar}(a) shows the $M_{4\pi}$ distribution under the $\pip\pim\pip\pim\gamma$ hypothesis with the requirement $\chisq_{4\pi} < 30$. The results are shown for the data and for the sum of the simulated ISR channels $\ep\en\to\Kp\Km\pip\pim\gamma$, $\ep\en\to\KS\Kpm\pimp\gamma$, $\ep\en\to\omega\pip\pim\gamma$ and the non-ISR continuum background. In Fig.~\ref{m4pi_babar}(b) these background contributions are shown individually. The $\chisq_{4\pi}$ requirement effectively eliminates the $\omega\pip\pim\gamma$ background: according to simulation, only 450 $\omega\pip\pim\gamma$ events remain after applying this restriction, in comparison to $260\,000$ signal events with a very similar $M_{4\pi}$ distribution, leading to a relative contribution of less than $\approx 0.2\%$. At very low invariant masses, $M_{4\pi}< 1.1\gevcc$, the background contributions from $\pip\pim\g\g$ and $\pip\pim\piz\g$ are large and not included in Fig.~\ref{m4pi_babar}. In this threshold region, the signal-to-background ratio is approximately 1:5. In the peak region, $1.1\gevcc < M_{4\pi} < 2.2\gevcc$, the background contamination is at the level of 3-4$\%$, dominated by $\ep\en\to\Kp\Km\pip\pim\g$ and $\ep\en\to\KS\Kpm\pimp\g$. At high $M_{4\pi}$, the background level rises to about 10\%, mainly due to the additional contribution of $\u\d\s$-continuum events.

A clear peak from $\jpsi\to\pip\pim\pip\pim$ is visible in the distributions of Fig~\ref{m4pi_babar}. In addition, the data exhibit a narrow peak in the \psitwos mass region, due to the decay $\psitwos\to\pipi \jpsi$ with $\jpsi\to\mup\mun$ and misidentified muons. The rejection of this background is described in subsection~\ref{subsec:bkglarge}. 

\subsection{Background in the peak region ($1.1\gevcc<M_{4\pi}<2.2\gevcc$)}
The dominant background in the peak region is from the ISR processes $\ep\en\to\Kp\Km\pip\pim\gamma$ and $\ep\en\to\KS\Kpm\pimp\gamma$. We utilize two different approaches to evaluate the background and use the difference between the methods to estimate the systematic uncertainty related to the background subtraction. The background subtraction itself is performed using method 2. 

\subsubsection*{Method 1: subtract simulated background}

For background subtraction via method 1, we use recent measurements of the most important background channels, $\Kp\Km\pip\pim\gamma$~\cite{2k2pi_new} and $\KS\Kpm\pimp\gamma$~\cite{2kpi}. 
We tune AFKQED according to these measurements and use the resulting predictions to evaluate the contributions of these channels to the $M_{4\pi}$ spectrum.

\subsubsection*{Method 2: background suppression}

This method is a hybrid approach. We impose specific requirements in order to suppress the $\Kp\Km\pip\pim\gamma$ and $\KS\Kpm\pimp\gamma$ backgrounds: the so-called $\Kp\Km\pip\pim\gamma$ and $\KS\Kpm\pimp\gamma$ vetoes. The remaining background is then subtracted according to method 1.

$\KS\Kpm\pimp\gamma$ events are vetoed if one of the tracks is identified as a \Kpm. The identification algorithm has an efficiency of 80-90\% per track and a $\pi$ mis-identification probability of 0.25\%. In addition we require the invariant mass formed from two of the three remaining tracks to lie within 35 \mevcc of the nominal $\KS$ mass~\cite{PDG10}. According to simulation, this method removes 74\% of the $\KS\Kpm\pimp\gamma$ background but only 1\% of signal.

$\Kp\Km\pip\pim\gamma$ events are rejected if two oppositely-charged tracks are identified as kaons and the requirement $\chisq_{2K2\pi}<10$ is fulfilled. The $\Kpm$ identification algorithm has an identification efficiency of 85-95\% per track and a $\pi$ mis-identification probability of 1\%. This requirement removes less than 0.1\% of the signal, but 55\% of the $\Kp\Km\pip\pim\gamma$ background.

Comparing the results of the two methods yields a systematic uncertainty of 1.0\% on the $\epem\to\pipi\pipi$ cross section in the peak region.

\subsection{Background at large invariant masses ($M_{4\pi}>2.2\gevcc$)}
\label{subsec:bkglarge}
From Fig.~\ref{m4pi_babar}(b) it is seen that the $\u\d\s$ continuum background is significant in the $M_{4\pi}>2.2\gevcc$ mass region. The largest contribution is from $\ep\en\to\pip\pim\pip\pim\piz$ events in which one of the photons from the $\piz\to\gamma\gamma$ decay is mistaken for the $\gamma_{ISR}$. The similar kinematic configuration of the $\u\d\s$-continuum events causes a peak at small values of $\chisq_{4\pi}$. We estimate this background by measuring the $\piz$ yield from a fit to the $\piz$ mass peak in the two-photon invariant mass distribution of the ISR photon candidate with any other photon candidate. We then scale the MC $\piz$ rate to give the same yield as observed in the data. The corresponding scaling factor is extracted with a relative uncertainty of about $22\%$. The normalized $\u\d\s$-continuum MC sample is used for background subtraction. The corresponding systematic uncertainty on the cross section measurement is 0.5\% in the peak region and rises up to 1.5\% for $M_{4\pi}>2.8\gevcc$.

\begin{figure}[h]
\epsfxsize=0.5\textwidth
\center{\epsfbox{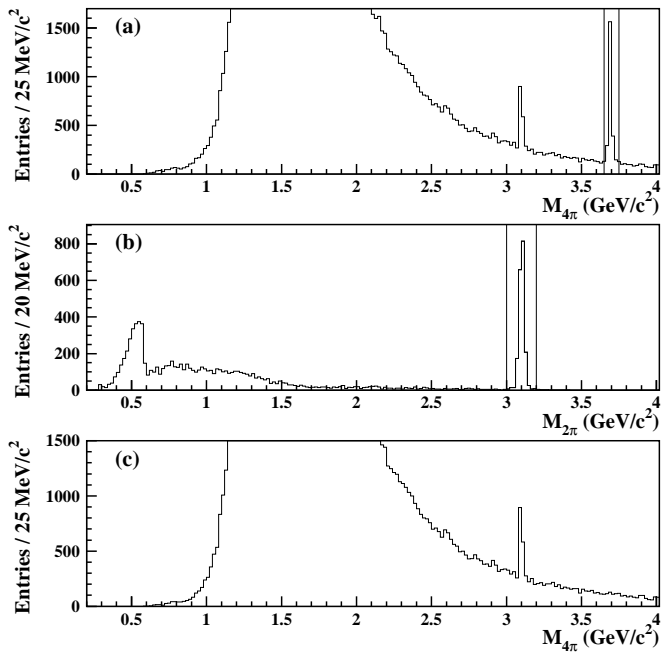}}
\caption{(a) Invariant mass distribution under the $\pip\pim\pip\pim$ hypothesis for the data; (b) $\pip\pim$ invariant mass distribution for events with $M_{4\pi}$ in the \psitwos mass region represented by the vertical lines in (a) (4 entries per event); (c) Same as part (a) after removing entries of the invariant $\pip\pim$-mass in the \jpsi mass region represented by the vertical lines in (b).}
\label{psi2s}
\end{figure}

There is an additional peaking background contribution in the charmonium region. Figure~\ref{psi2s}(a) shows the invariant mass distribution under the $\pip\pim\pip\pim$ hypothesis for the data. Peaks corresponding to the \jpsi and \psitwos resonances are clearly seen. In Fig.~\ref{psi2s}(b) the invariant $\pipi$ mass $M_{2\pi}$ distribution for events within the \psitwos ($3.65\gevcc<M_{4\pi}<3.75\gevcc$) mass region are plotted. The distribution contains four entries per event. The \jpsi peak is clearly visible. The $\jpsi\to\pip\pim$ branching fraction is about a factor 400 smaller than that of $\jpsi\to\mup\mun$~\cite{PDG10}. No PID selector is used in the sample selection. In addition, the kinematic fit does not suppress $\pip\pim\mup\mun$ events, because the mass difference between muons and pions is only 34\mevcc, which is negligible due to the fact that they are highly relativistic, being emitted from the \jpsi. Thus, the observed peak is likely dominated by $\psitwos\to\pip\pim\mup\mun$ decays. To remove this background, we reject events that have $M_{4\pi}$ in the \psitwos region and an $M_{2\pi}$ value consistent with the \jpsi. The corresponding $M_{4\pi}$ distribution after the \psitwos veto is shown in Fig.~\ref{psi2s}(c).

\subsection{Background at small invariant masses ($M_{4\pi}<1.1\gevcc$)}\label{2pi2el}
Background in the threshold region, i.e., below 1.1\gevcc, is dominated by two processes: \g conversion of a real photon in the detector material, and conversion of a virtual photon at the primary interaction vertex (Dalitz conversion). In $\ep\en\to\pip\pim\piz\gamma_{\mathrm{ISR}}$ events, one of the photons from the $\piz$ decay can convert in the detector material into an $\ep\en$ pair, and both the $\ep$ and the $\en$ can be misidentified as pions. Moreover $\ep\en\to\pip\pim\gamma_{ISR}\gamma$ events in which the non-ISR photon converts into an $\ep\en$ pair contribute a similar background. The second source of background in this mass region is $\ep\en\to\pip\pim\gamma_{ISR}\ep\en$ events in which the $\ep\en$ pair arises from a Dalitz conversion process. Two methods are used to remove these background channels.

\subsubsection*{Method 1: Vetoes for Dalitz conversion and conversion in the detector}

The first method vetoes events with either a primary vertex probability less than $10^{-8}$ or with two identified electrons. Electrons are identified using an algorithm with an efficiency of 99\% and a pion mis-identification rate of 5-10\% depending on the transverse momentum of the track. The first requirement is not sufficient to reject background events with highly-energetic electrons or positrons, which have a vertex probability similar to non-conversion events. MC study shows that the combination of both requirements yields a background rejection larger than 99\% while removing less than 6\% of signal.
\subsubsection*{Method 2: Pion identification}
The second method requires all four tracks to be identified as pions, using a selector with an efficiency of 97-99\% and an electron mis-identification probability of 5-7\% depending on the transverse momentum of the track. There is, however, a difference in efficiency between the data and MC simulation of approximately $0.5-1\%$ per track ($2-4\%$ shift per event). Therefore this selector is not used in the peak region, where the background contribution is very low.\\

\begin{figure}[h]
\begin{center}
\includegraphics[clip, width=0.45\textwidth]{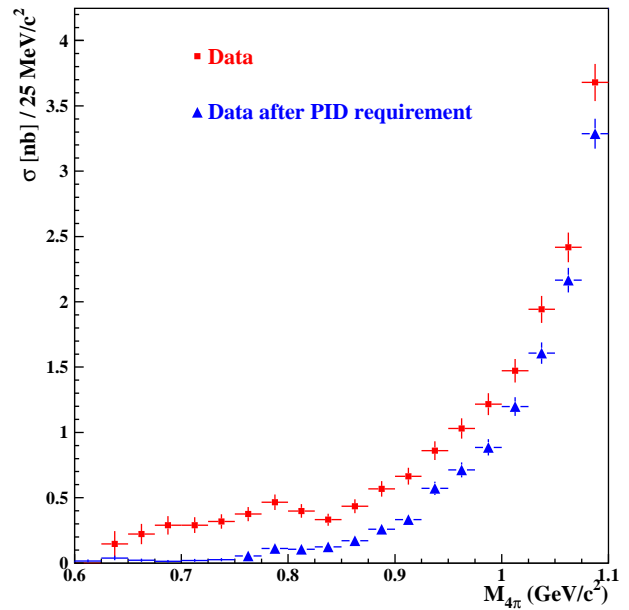}
\end{center}
\caption{$\pip\pim\pip\pim\gamma$ cross section in the threshold region for the data (squares) and for the data after requiring each track to satisfy a dedicated pion selection algorithm (triangles), corresponding to the method 2 selection criterion. The indicated uncertainties are statistical.}
\label{xs_small}
\end{figure}

Both methods remove a large fraction of the conversions and yield results that are consistent with each other. We present our primary results using the pion selection method, as the systematic uncertainties on its inputs are better understood. Comparing the cross section results using the two methods leads to a systematic uncertainty estimate of 3\%. It should be noted that the $\rho(770)^0$ peak is strongly suppressed, but is still visible as a shoulder after background subtraction, as shown in Fig.~\ref{xs_small}.

\subsection{Background subtraction summary}

\begin{figure}[t]
\begin{center}
\includegraphics[clip, width=0.5\textwidth]{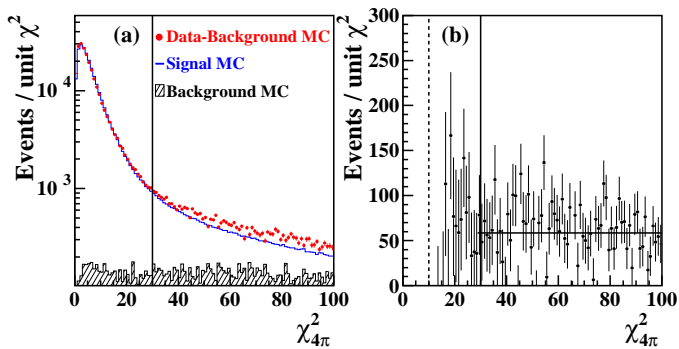}
\end{center}
\caption{(a) The $\chisq$ distribution for data after background subtraction (points), signal MC (histogram) and the sum of simulated background channels (hatched histogram). (b) The difference between data with background subtraction and signal MC. The dashed line indicates the boundary with the $0<\chisq_{4\pi}<10$ region used for normalization.}
\label{4pi_chisq_all}
\end{figure}

Figure~\ref{4pi_chisq_all}(a) shows the $\chisq_{4\pi}$ distribution both for the data after background subtraction and the signal simulation. Compared to Fig.~\ref{4pi_babar}(a), the difference between the data and the signal simulation is reduced. There is still some remaining background at larger $\chisq_{4\pi}$ values as can be seen in Fig.~\ref{4pi_chisq_all}(b). This is consistent with the uncertainties in the background-subtraction methods as previously described. To be conservative, we assume that the remaining background is uniform as a function of $\chisq_{4\pi}$. In this case, we find that the associated uncertainty on the cross section is less than 0.4\% for $M_{4\pi} < 2.8\gevcc$ and 4.0\% for $2.8\gevcc < M_{4\pi} < 4.5\gevcc$.

\section{Acceptance and efficiencies}
\label{sec:acc}

\begin{figure}[h]
\centerline
{
\epsfxsize=0.5\textwidth
\epsfbox{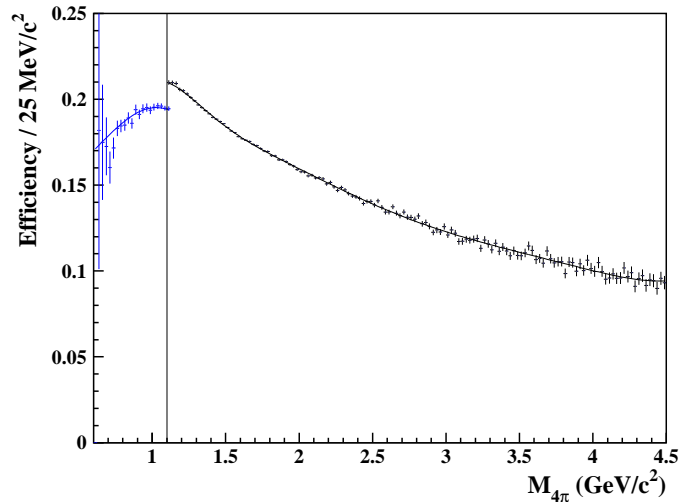}
}
\caption{The efficiency as a function of the $\pip\pim\pip\pim$ invariant mass. The vertical line at $M_{4\pi}=1.1\gevcc$ separates the regions with and without a pion PID requirement for the reconstructed tracks. The curves on either side of the vertical line are fits based on the sum of a Gaussian and a polynomial.}
\label{mc_acc}
\end{figure}

The full chain of experimental requirements is applied to the signal MC sample. We define the ratio of the number of selected events divided by the number of events without applying any requirements as the global efficiency. Figure~\ref{mc_acc} shows the global efficiency as a function of $M_{4\pi}$ determined with the simulation. The decrease of efficiency in the threshold region, which corresponds to the highest energy ISR photons, is due to the fact that the four tracks of the hadronic system recoil in a narrow cone opposite to the direction of the ISR photon. Since ISR photons are preferentially emitted at small polar angles, the efficiency for detecting all four tracks within the fiducial volume of the detector decreases accordingly. The decrease in the global efficiency at large values of $M_{4\pi}$ can be explained with similar arguments, because the opening angle of the hadronic system increases with decreasing ISR photon energy, increasing the probability of losing one of the four tracks. The discontinuity at $M_{4\pi}=1.1\gevcc$ is due to the pion PID requirement for the tracks at low invariant masses.

As mentioned in section~\ref{sec:babar}, it is assumed that the $\pip\pim\pip\pim\gamma$ hadronic final state arises from the decay of various intermediate resonances such as $a_1(1260)\pi$. The relative contributions of intermediate states are discussed in section~\ref{sec:inv}. Different intermediate states might exhibit different results for the angular distributions of final-state particles. The limited knowledge of the hadronic substructure hence corresponds to a systematic uncertainty in the evaluation of the global efficiency. This effect is estimated through comparison of the predictions from AFKQED and a new version of PHOKHARA~\cite{phok}, which contain different intermediate resonances. The effect is observed to be negligible.  

\subsection{Photon efficiency}

\begin{figure}[h]
\begin{center}
\epsfxsize=0.5\textwidth
\epsfbox{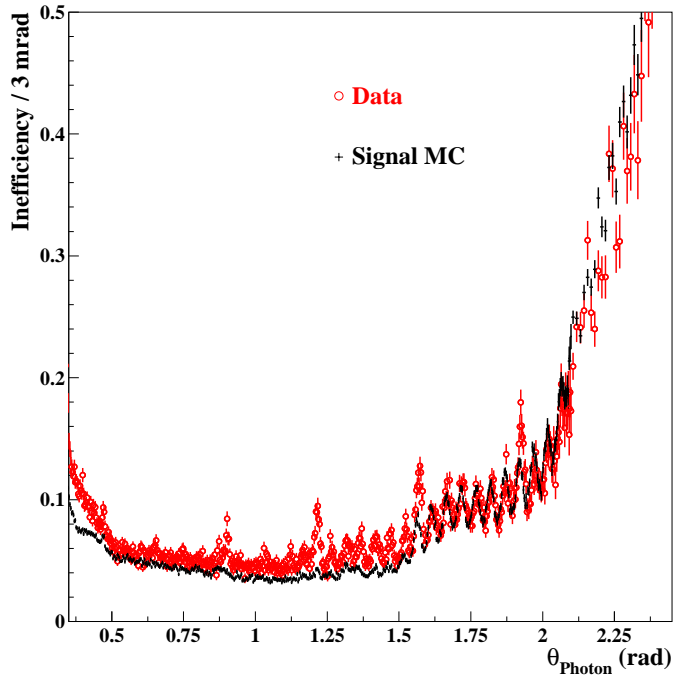}
\end{center}
\caption{\label{ineff_ISR} Photon inefficiency as a function of the ISR photon polar angle for the data (circles) and signal MC (crosses). The binsize is increased to 6\mrad for $\theta_\text{Photon}>2.1\rad$.
}
\end{figure}

A dedicated study is performed to determine the photon efficiency in data and simulation. Detector inefficiencies due to inactive material between crystals, non-functioning crystals, and conversions in inner detector structures are investigated. This study is performed with a $\mumu\g_{ISR}$ sample, in which an identified ISR photon is not required in the event selection. A kinematic fit with one constraint (1C) based on the kinematic information of the charged tracks is performed in order to calculate the energy and direction of the photon. This fit prediction is compared to the measured photon information to extract the inefficiency of the photon reconstruction. In Fig.~\ref{ineff_ISR} the photon inefficiency is shown as a function of the polar angle of the photon for the data and simulation. The inefficiency curve is not smooth due to the effect of the inactive regions between EMC crystal rings. Some gaps between rings, which are visible in the data as peaks of high inefficiency, are not properly simulated, especially in the forward region of the detector. As a function of $M_{4\pi}$, we observe a uniform inefficiency difference between the data and the simulation with an average value of $\Delta\epsilon_{\gamma} = (1.34\pm 0.03_{stat}\pm 0.37_{syst})\%$.

\subsection{Tracking efficiency}

Track reconstruction and effects like nuclear interactions and energy loss of tracks traversing the detector volume are not simulated perfectly. As a consequence, the tracking efficiency is slightly different for the data and MC. This difference is investigated with a dedicated study of ISR $\ep\en\to\pipi\pipi$ events with one missing track. The missing momentum and direction of the lost pion is calculated using a constrained kinematic fit with one remaining constraint (1C). The angular and momentum dependent distribution of the inefficient events is extracted.

Small differences in tracking efficiency between the data and simulation are due to an imperfect description of track loss when tracks overlap in azimuth. The average difference between the data and MC is uniform as a function of the transverse momentum and the polar angle of the tracks and is determined on average to be $\Delta\epsilon_{trk} = (0.75 \pm 0.05_{stat} \pm 0.34_{syst})\%$.

\section{Cross section}
\label{sec:xs}
As described in section~\ref{sec:Introduction}, the non-radiative cross section is related to the radiative cross section according to:
\begin{equation}
\sigma_{4\pi}(M_{4\pi}) = \frac{d\sigma_{4\pi,\g} (M_{4\pi})}{dM_{4\pi}}\cdot \frac{s}{2M_{4\pi}}\cdot \frac{1}{W(s,x,C^{\ast})}.
\label{equ:xs}
\end{equation}
The radiative cross section is:
\begin{equation}
\frac{d\sigma_{4\pi,\g} (M_{4\pi})}{dM_{4\pi}} = \frac{dN_{4\pi,\g} (M_{4\pi})}{dM_{4\pi}}\cdot\frac{1}{{\cal L}_{tot}\cdot\epsilon\cdot(1+\delta_{rad,FSR})}
\label{equ:radxs}
\end{equation}
where $dN_{4\pi,\g}$ is the number of selected events, $\epsilon$ the global efficiency corrected for tracking and photon efficiency differences between the data and MC, and $\delta_{rad,FSR}$ the radiative corrections including LO- and NLO-FSR effects. The MC event generator interfaced with the detector simulation is based on a modified version of the EVA code~\cite{kuehn2}. It contains collinear NLO-ISR corrections based on the structure function technique, as well as FSR corrections based on PHOTOS~\cite{PHOTOS}. Radiative corrections due to NLO-ISR on the radiator function obtained with this event generator are compared to the PHOKHARA generator~\cite{phok}, which includes the full NLO-ISR corrections. 
With our selection, a $(1.0\pm0.2)\%$ difference of the radiator function between the two generators is observed. We apply a correction to account for this difference.

\begin{figure}[h]
\epsfxsize=0.5\textwidth
\epsfbox{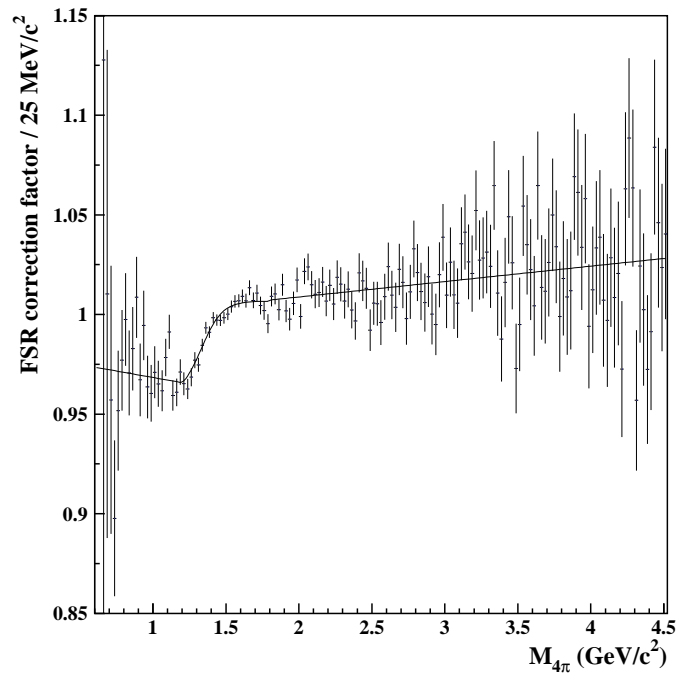}
\caption
{Ratio of the simulated radiative cross section with only the ISR contribution to the cross section with both ISR and FSR.}
\label{fsr_effect}
\end{figure}

Figure~\ref{fsr_effect} shows the ratio of the simulated ISR radiative cross section and the cross section including additional FSR (PHOTOS). The FSR leads to a shift of events towards lower invariant masses in the radiative cross section, due to the fact that the measured invariant mass is smaller than the effective $E_{CM}$ for events with FSR.

\begin{figure}[h]
\epsfxsize=0.5\textwidth
\epsfbox{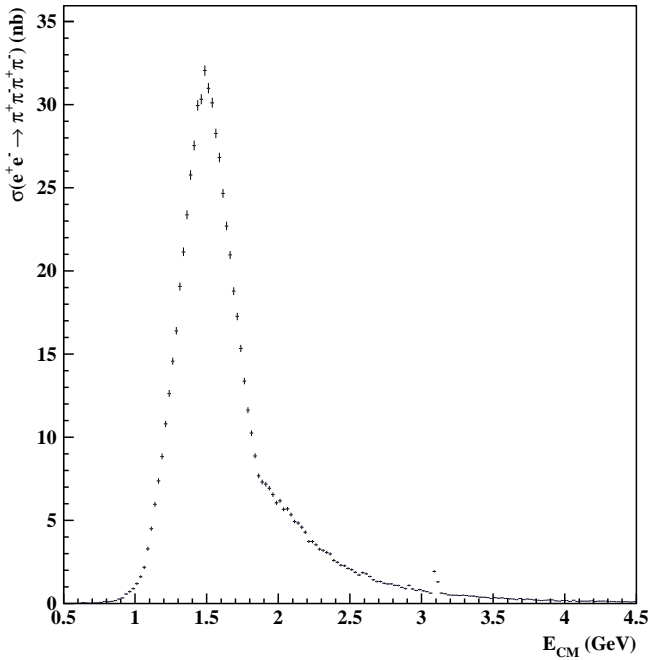}
\caption
{The $E_{CM}$ dependence of the dressed $\ep\en\to\pipi\pipi$ cross section measured from the ISR data. The uncertainties are statistical.}
\label{4pi_ee_babar}
\end{figure}

After applying all radiative corrections and accounting for the relevant differences in efficiencies between the data and simulation, we apply eq.~(\ref{equ:xs}) and extract the non-radiative $\ep\en\to\pipi\pipi$ cross section. The result is shown in Fig.~\ref{4pi_ee_babar}. 
The measured cross section includes the contributions of vacuum polarization. 
The cross section used in the dispersion integral for $a_{\mu}$ does not include vacuum polarization and thus we need to apply a correction.
We define the {\em undressed} cross section $\sigma_{4\pi}^{ud}(E_{CM})$ by correcting the measured or {\em dressed} cross section $\sigma_{4\pi}^{d}(E_{CM})$ for vacuum polarization effects according to:
\begin{equation}
\sigma_{4\pi}^{ud}(E_{CM})=\sigma_{4\pi}^{d}(E_{CM})\cdot {\left(\frac{\alpha(0)}{\alpha(E_{CM})}\right)}^2 = \frac{\sigma_{4\pi}^{d}(E_{CM})}{\delta_{vac}(E_{CM})}
\label{equ:undress}
\end{equation}
where $\alpha(E_{CM})$ is the electroweak coupling strength at $E_{CM}$. The correction due to vacuum polarization $\delta_{vac}(E_{CM})$, which can be found in Ref.~\cite{vacpol}, is applied.  Our results for the dressed and the undressed cross sections are presented in Table~\ref{xs_sum}.

\input {xs_table}

\subsection{Systematic corrections and uncertainties}
Table~\ref{tab2} presents the complete list of corrections and systematic uncertainties that are included in the dressed cross section. The uncertainties associated with background subtraction are discussed in section~\ref{sec:bkg}. The 3.0\% tracking efficiency difference between data and MC has an uncertainty of 1.4\%. The photon efficiency correction is $1.3\pm0.4\%$. The total luminosity is measured with a precision of 1.0\%. A  $1.0\pm0.2\%$ difference is observed between the radiator functions computed with AFKQED and PHOKHARA. The effect of additional FSR is estimated using PHOTOS, resulting in the correction shown in Fig.~\ref{fsr_effect} and a systematic uncertainty of 0.5\%. The requirement $\chisq_{4\pi}<30$ leads to a systematic uncertainty of 0.3\%. The uncertainty on the global efficiency is estimated to be 1.0\% in the central region, increasing to 10\% in the low mass region $M_{4\pi}<1.1\gevcc$ due to an observed efficiency decrease of up to 10\%. A conservative uncertainty of 10\% to account for the total acceptance decrease in this region is also applied.

\begin{table*}
\caption{\label{tab2}Summary of systematic corrections and uncertainties in per centage.}
\begin{longtable*} {c c c c c}
\hline
\hline
\noalign{\vskip1pt}
 $M_{4\pi}$ & $< 1.1\gevcc$ & $<2.8\gevcc$ &  $<4.0\gevcc$ & $<4.5\gevcc$\\
\hline
\noalign{\vskip1pt}
$\Kp\Km\pipi\gamma$, $\KS\Kpm\pimp\gamma$ & $\pm$ 1.0 & $\pm$ 1.0 & $\pm$ 3.0 & $\pm$ 7.0\\
continuum bkg & - & $\pm$ 0.5 & $\pm$ 1.0 & $\pm$ 1.5\\
$\pipi e^+ e^- \g$ & $\pm$ 3.0 & - & - & - \\
additional bkg & $\pm$ 0.4  & $\pm$ 0.4  & $\pm$ 4.0  & $\pm$ 4.0 \\
\hline
tracking efficiency & +3.0 $\pm$ 1.4 & +3.0 $\pm$ 1.4 & +3.0 $\pm$ 1.4 & +3.0 $\pm$ 1.4\\
photon efficiency & +1.3 $\pm$ 0.4 & +1.3 $\pm$ 0.4  & +1.3 $\pm$ 0.4  & +1.3 $\pm$ 0.4  \\
${\cal L}$ & $\pm$ 1.0 & $\pm$ 1.0  & $\pm$ 1.0  & $\pm$ 1.0 \\
AFK-PHOK-difference & $-1.0$ $\pm$ 0.2 & $-1.0$ $\pm$ 0.2 & $-1.0$ $\pm$ 0.2  & $-1.0$ $\pm$ 0.2  \\
FSR corrections & $\pm$ 0.5 & $\pm$ 0.2  & $\pm$ 0.1  & $\pm$ 0.1 \\
$\chisq_{4\pi}<30$ & $\pm$ 0.3 & $\pm$ 0.3  & $\pm$ 0.3  & $\pm$ 0.3  \\
global efficiency & $\pm$ 10.0 & $\pm$ 1.0 & $\pm$ 1.0 & $\pm$ 1.0 \\
\hline
sum & $\pm$ 10.7 & $\pm$ 2.4 & $\pm$ 5.5 & $\pm$ 8.5 \\
\hline
\hline
\end{longtable*}
\end{table*}

Assuming no correlation between the various contributions to the systematic uncertainty of the cross section, its total is found to be 10.7\% for $M_{4\pi}<1.1\gevcc$, 2.4\% for $1.1\gevcc<M_{4\pi}<2.8\gevcc$, 5.5\% for $2.8\gevcc<M_{4\pi}<4.0\gevcc$ and 8.5\% for higher invariant masses.
Individual contributions to the systematic uncertainties contribute in a correlated way on the whole mass range, with the exception of the global analysis efficiency, for which it does not. Therefore for $M_{4\pi}>1.1\gevcc$ a 100$\%$ correlation can be assumed, while for $M_{4\pi}<1.1\gevcc$ where the global efficiency dominates, it can be assumed to be uncorrelated.

\subsection{Comparison with the existing $\ep\en$ data}

\begin{figure}[htb]
\epsfxsize=0.5\textwidth
\center{\epsfbox{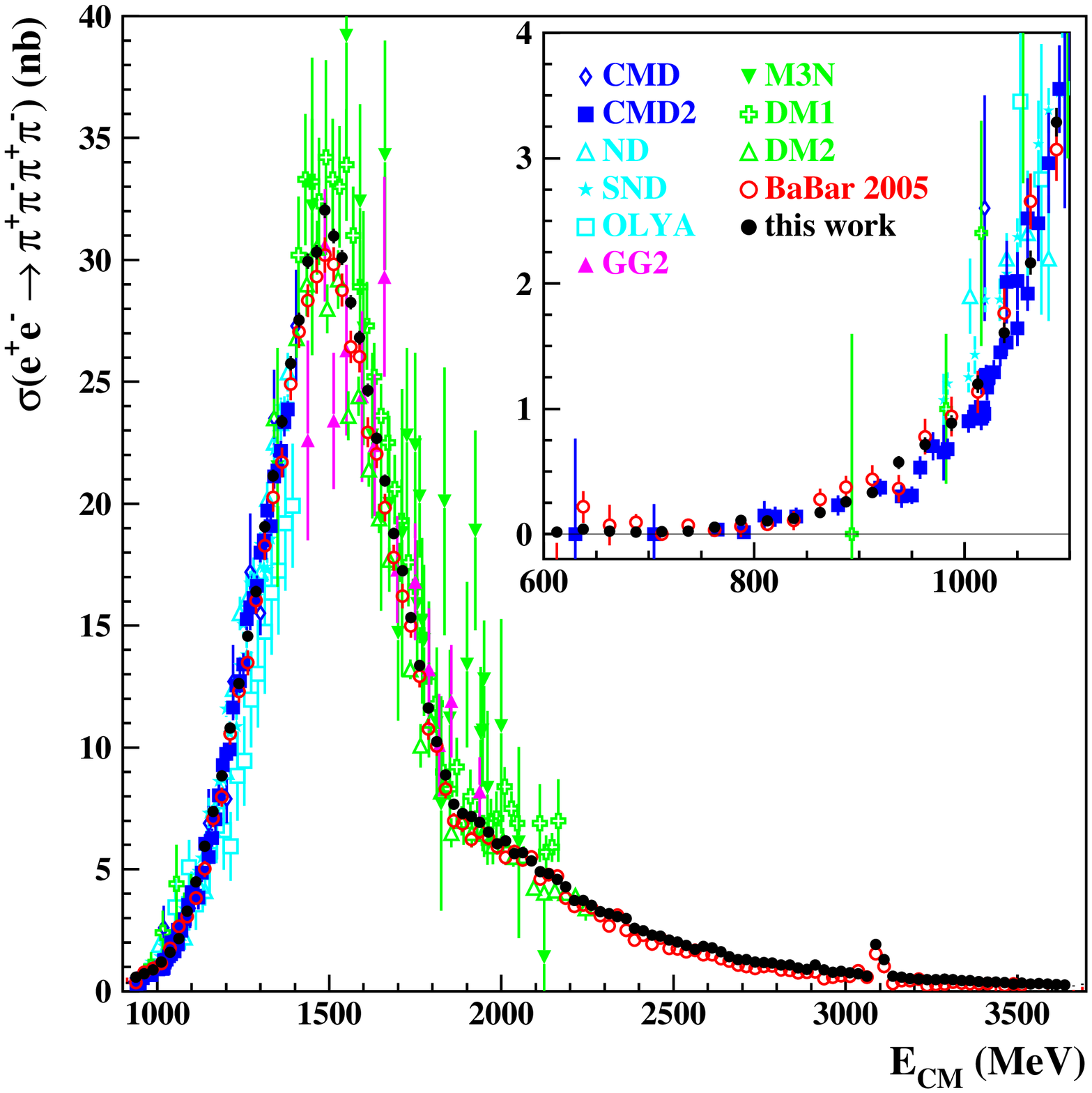}}
\caption {The $\ep\en\to\pip\pim\pip\pim$ cross section in comparison to other experiments. Only statistical uncertainties are shown.}
\label{comp_worlddata}
\end{figure}

In Fig.~\ref{comp_worlddata} the extracted non-radiative $\sigma(e^+e^-\to\pip\pim\pip\pim)$ cross section is shown, in comparison with the previous \babar\ result~\cite{4piold} and the results from fixed-energy $\ep\en$ experiments. Our results agree within the uncertainties with our previous measurement, which they supersede. Our results are consistent with and higher in precision than the direct $\ep\en$ cross section measurements made at VEPP-2M by OLYA~\cite{4pi_olya}, ND~\cite{4pi_nd}, SND~\cite{4pi_snd}, CMD~\cite{4pi_cmdexp}, and CMD-2~\cite{4pi_cmdnew,4pi_cmdold_2,4pi_cmdold}, at DCI by M3N~\cite{4pi_m3n}, DM1~\cite{4pi_dm1}, and DM2~\cite{4pi_dm2} and at Adone by GG2~\cite{4pi_gg2}.

\subsection{Influence on the prediction of $a_{\mu}$}
Using the result for the $\ep\en\to\pipi\pipi$ cross section obtained in the present study, we compute the contribution of this channel to the anomalous magnetic moment of the muon $a_{\mu}$ via a dispersion relation using the HVPTool program~\cite{hvptools} in the energy region $0.6\gev<E_{CM}<1.8\gev$. We find:
\begin{equation}
a_\mu^{had}(\pipi\pipi)=(13.64\pm0.03_{stat}\pm0.36_{syst})\times10^{-10}.
\end{equation}
Our result is more precise than the current world average for this quantity: $(13.35 \pm 0.10_{stat} \pm 0.52_{syst})\times10^{-10}$~\cite{davg-2}, where the first uncertainty is statistical and the second systematic.

\section{Invariant masses and charmonium branching ratios}
\label{sec:inv}
\begin{figure}[tbh]
\centerline
{
\epsfxsize=0.5\textwidth
\epsfbox{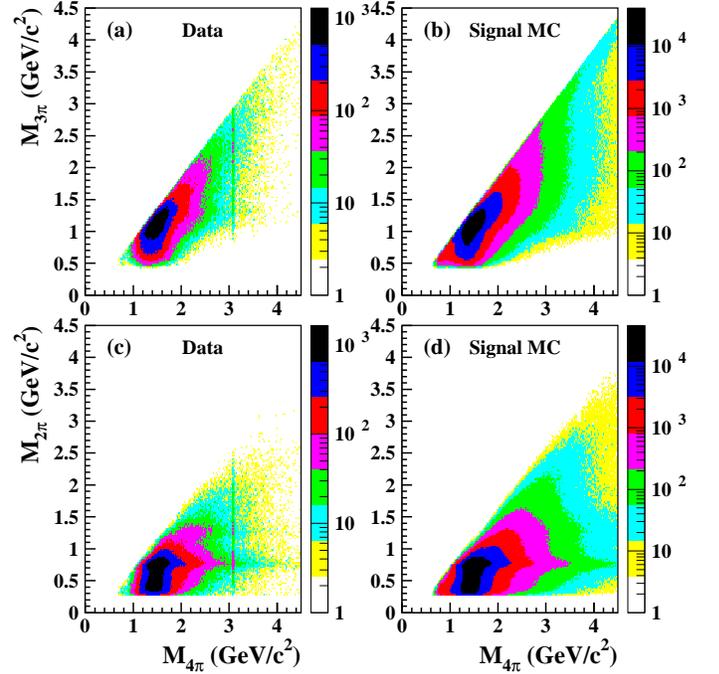}
}
\caption{Invariant $\pipi\pipm$ and $\pipi$ mass combinations vs. invariant $\pipi\pipi$ mass for the data without background subtraction (left) and signal MC (right).
}
\label{4pi_2-3pi}
\end{figure}

Different invariant mass combinations have been studied in the data and MC simulation to search for states not included in the MC model. In the following, we present a general qualitative search for these hadronic structures. We then consider a more detailed study of the \jpsi and \psitwos background subtraction and efficiency corrections. Finally, we determine the branching fractions $\BR_{\jpsi\to \pip\pim\pip\pim}$ and $\BR_{\psitwos\to \jpsi \pip\pim}$ and perform a scan for additional resonances at high invariant masses.
 
\subsection{Substructures}

The scatter plots in Fig.~\ref{4pi_2-3pi} display distributions of the invariant $\pipi\pipm$ and $\pipi$ masses versus the invariant $\pipi\pipi$ mass for the data and MC. The $\rho(770)^0$ band is clearly visible in the $\pipi$ mass distribution of the data and MC. In general, good agreement is seen except for the \jpsi decay, which is not simulated.

In a more detailed study the $\pipi\pipi$ mass spectrum is divided into five intervals:
\begin{enumerate}
\item{1.0-1.4\gevcc:} low mass region
\item{1.4-1.8\gevcc:} peak region of the cross section
\item{1.8-2.3\gevcc:} high mass shoulder
\item{2.3-3.0\gevcc}
\item{3.0-4.5\gevcc:} without the narrow region around \jpsi
\end{enumerate}

\begin{figure*}[tbh]
\centerline
{
\epsfxsize=.95\textwidth
\epsfbox{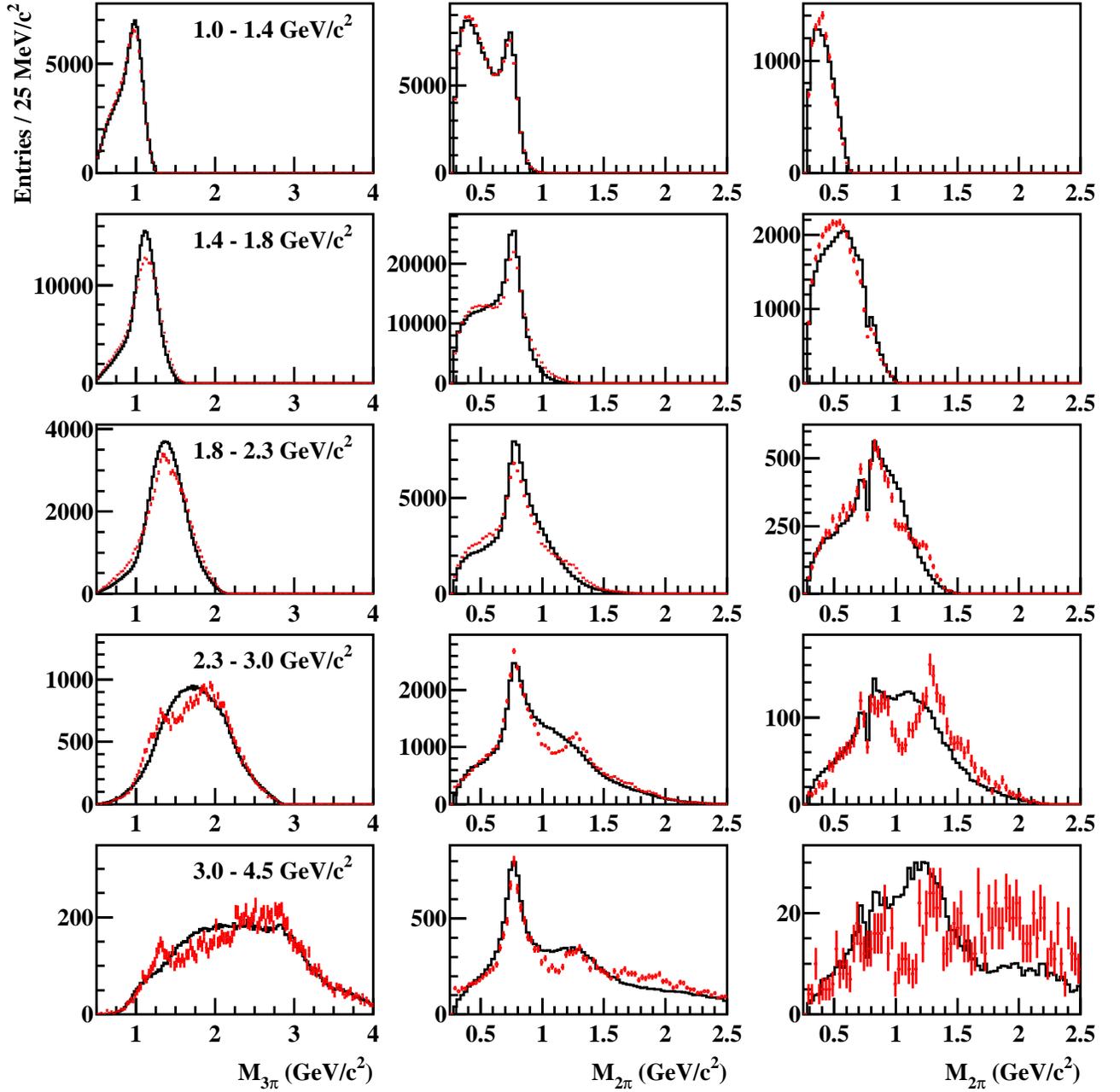}
}
\caption{Left: invariant $\pipi\pipm$ mass distributions (4 entries per event) for different regions in $M_{4\pi}$ for the data without background subtraction (points) and signal simulation (histogram); Middle: $\pipi$ mass distributions (4 entries per event); Right: $\pipi$ mass distribution with another $\pipi$ mass in the $\rho(770)^0$ mass region. }
\label{2pi_3pi_pro}
\end{figure*}

Figure~\ref{2pi_3pi_pro} shows the one-dimensional distributions from the five regions for the two- and three-pion invariant masses in comparison with MC~\cite{kuehn2}.

The $\pipi\pipm$ invariant mass distribution is shown in the leftmost column of Fig.~\ref{2pi_3pi_pro}. In the low mass region, $1.0\gevcc<M_{4\pi}<1.4\gevcc$, there is not enough energy to allow production of the $a_{1}(1260)^\pm$. At higher $M_{4\pi}$, the contribution of the $a_{1}(1260)^\pm$ becomes visible. It is observed as a peaking structure with mass and width $M_{3\pi}\approx 1300\mevcc$ and $\Gamma\approx 200\mev$. In comparison, the average mass value in the PDG~\cite{PDG10} is $1.230\pm0.040\gevcc$, with results from individual experiments that vary between 1.04 and 1.33\gevcc. The corresponding width varies between 250 and 600\mev. In our simulation, the parameters $M=1.33\gevcc$ and $\Gamma=570\mev$ are used, which are determined from a combined analysis of the CLEO and CMD-2 data~\cite{Bondar}. Our results seem to favor a lower $a_{1}(1260)^\pm$ mass and a smaller width.

In the $\pipi$ invariant mass distributions shown in the middle column of Fig.~\ref{2pi_3pi_pro}, four entries are present per event. At low $4\pi$ mass and in the peak region only a single resonance, the $\rho(770)^0$, is observed. At larger $4\pi$ mass, a second peaking structure appears at $M_{2\pi}\approx 1270\mevcc$, which most likely corresponds to the $f_2(1270)$. This resonance is not simulated by our MC. It is observed that over the entire $4\pi$ mass range, approximately 25\% of the entries are in the $\rho(770)^0$ peak. $\rho(770)^0\rho(770)^0$ production is not allowed due to C-parity conservation, leading to the conclusion that in each event one $\rho(770)^0$ meson is present.

To investigate the possible presence of the $f_{2}(1270)\rho(770)^0$ final state, the $\pip\pim$ combination is plotted for the case that there is another $\pip\pim$ combination within $\pm25\mevcc$ of the $\rho(770)^0$ mass, $745\mevcc < M_{2\pi}<795\mevcc$. The results are shown in the rightmost column of Fig.~\ref{2pi_3pi_pro}. An artificial dip at $M_{2\pi}\approx 770\mevcc$ due to the selection of the $\rho(770)^0$ is observed. The $f_{2}(1270)$ resonance is visible as a shoulder in the $1.8\gevcc<M_{4\pi}<2.3\gevcc$ mass region. It is even more prominent in the $2.3\gevcc<M_{4\pi}<4.5\gevcc$ region, where the energy is large enough to allow direct production of $f_{2}(1270)\rho(770)^0$. A sharp falloff in the $M_{2\pi}$ spectrum just below 1\gevcc is visible in the  $1.8\gevcc<M_{4\pi}<2.3\gevcc$ region. This might be due to interference with the $f_0(980)$ final state. A partial wave analysis, which is beyond the scope of this paper, would be necessary to determine the structure of the individual intermediate states.
A qualitative comparison with the MC model of Ref.~\cite{phok} indicates a somewhat better agreement with the data than that shown in the rightmost column of Fig.~\ref{2pi_3pi_pro}, apart from an overestimate of the contribution of the $f_0(1300)\rho(770)^0$ final state.

\subsection{\jpsi and \psitwos}\label{resolution}
\begin{figure}[h]
\epsfxsize=0.235\textwidth
\epsfbox{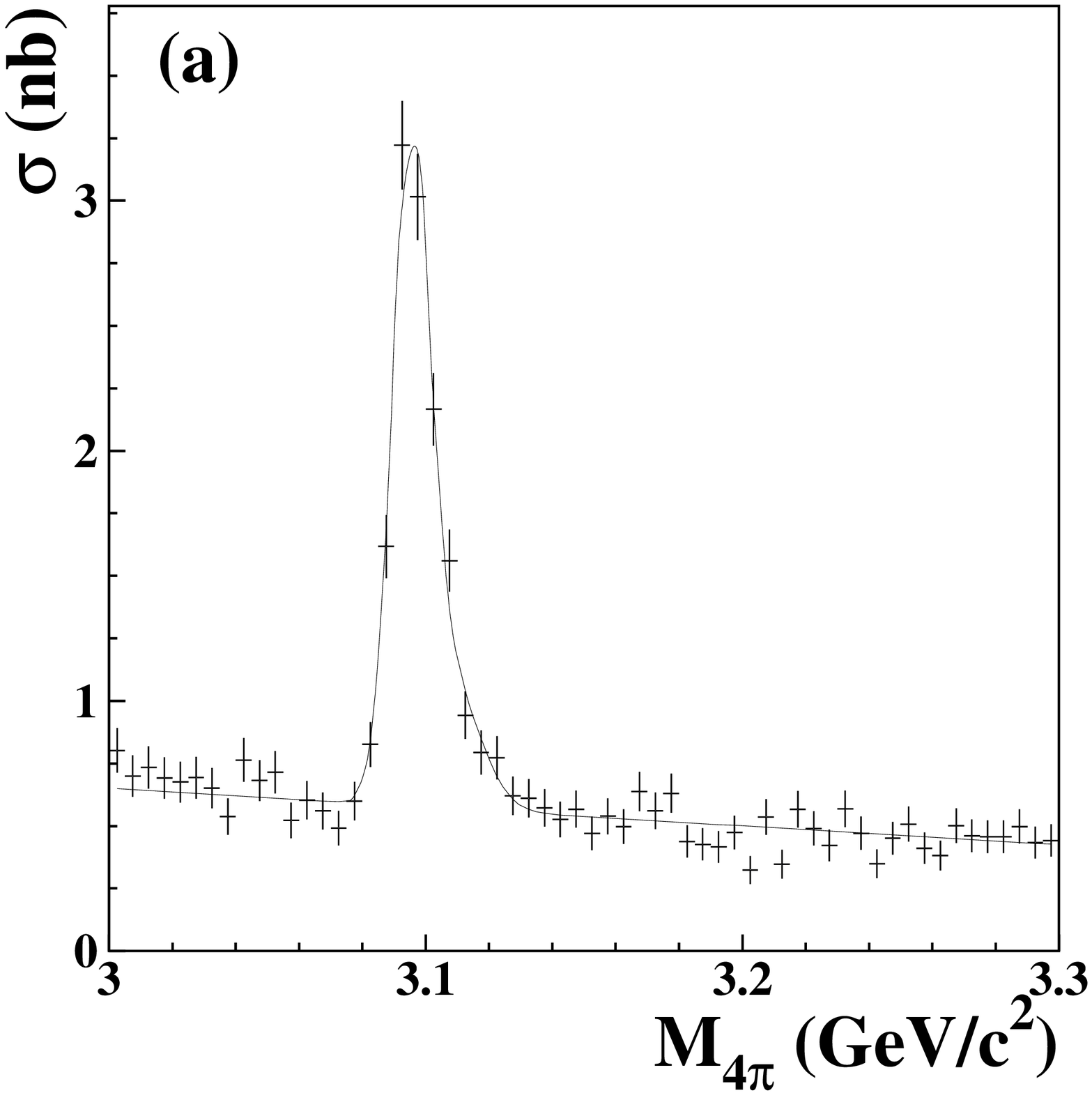}
\epsfxsize=0.235\textwidth
\epsfbox{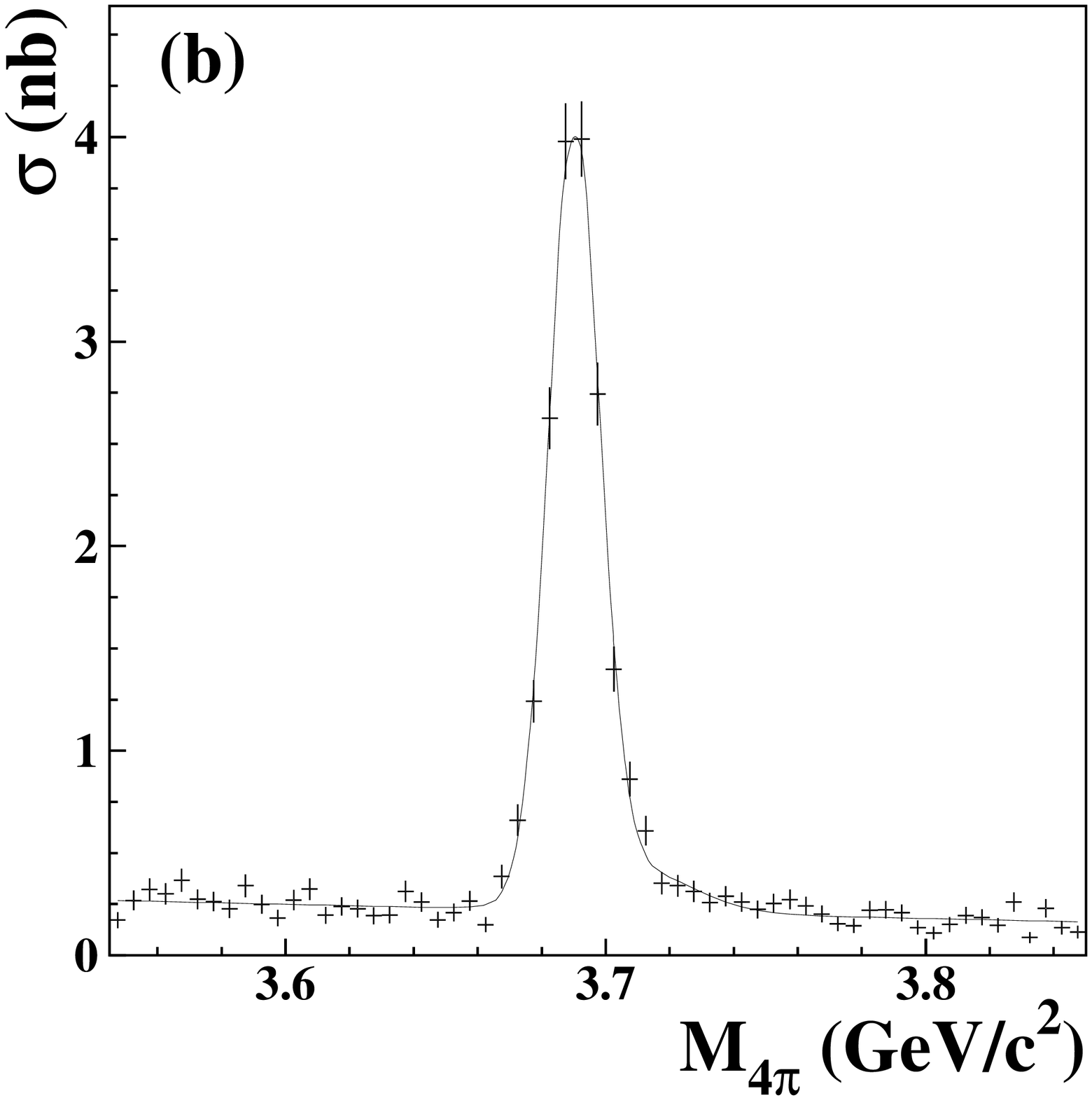}
\caption {(a) Invariant 4$\pi$ mass $M_{4\pi}$ distribution for the data around the $\jpsi$ peak with a fit, which consists of a sum of two Gaussian and a linear distribution, describing the peak and the non-resonant $\pipi\pipi$ contribution, respectively. (b) Invariant $\pipi\pipi$ mass $M_{4\pi}$ distribution, assuming all four particles are pions, for the data around the \psitwos peak with a fit, which consists of a sum of two Gaussian and a linear distribution, describing the peak and the non-resonant $\pipi\pipi$ contribution, respectively.}
\label{psi2s_fit}
\end{figure}

Figure~\ref{psi2s_fit}(a) displays the $\sigma(\ep\en\to\pipi\pipi)$ cross section as a function of the $4\pi$ mass, in the vicinity of the $\jpsi$ meson. The measured width of the \jpsi ($\approx15\mev$) is dominated by the track momentum resolution, the intrinsic width being $\Gamma_{\jpsi}=93\kev$~\cite{PDG10}. The small tail towards higher masses is mostly from extra radiation, which is assigned to the hadronic system by the fit. We describe the $\jpsi$ peak and the non-resonant $\pip\pim\pip\pim$ contribution with the sum of two Gaussians and a linear term, respectively. This allows the integrated partial cross section $\sigma^{\jpsi}_{int}=\int^{\infty}_{0}dM_{4\pi}\sigma^{\jpsi}(M_{4\pi})$ and the electronic width of $\jpsi\to\pipi\pipi$ to be extracted:

\begin{eqnarray}                                                        
&& \hspace{0cm}\BR_{\jpsi\to \pip\pim\pip\pim}\cdot \sigma^{\jpsi}_{int}\nonumber\\                                          
&=&\frac{N(\jpsi\to\pip\pim\pip\pim)}{d{\cal L}/dE\cdot\epsilon_{MC}}\nonumber \\                                             
&=& (48.9\pm 2.1_{stat}\pm1.0_{syst})\mevcc\nb                                                                                
\end{eqnarray}                                                                                                                
and

\begin{eqnarray}
&&\hspace{0.0cm}\BR_{\jpsi\to \pip\pim\pip\pim}\cdot \Gamma^{\jpsi}_{ee}\nonumber \\
&=& \frac{N(\jpsi\to\pip\pim\pip\pim)\cdot M^2_{\jpsi}}{6\pi^2\cdot d{\cal L}/dE\cdot\epsilon_{MC}\cdot C}\nonumber \\
&=& (20.4\pm0.9_{stat}\pm0.4_{syst})\ev.
\end{eqnarray}

For the above, the value $M_{\jpsi}=3096.92\pm0.01\mevcc$~\cite{PDG10} and the conversion constant $C=3.8938\times 10^{11} \mev^2 \nb$~\cite{PDG10} are used. The statistical uncertainty corresponds to the fit uncertainty on the area under the two Gaussian distributions, which is a fit parameter. A systematic uncertainty of 3$\%$ covers the systematic effects related to the luminosity and efficiencies. Contributions from background that peak at the \jpsi mass are negligible.
Using the electronic width $\Gamma_{ee}^{\jpsi}=5.55\pm0.14\kev$, we determine the $\jpsi\to \pip\pim\pip\pim$ branching fraction to be:

\begin{eqnarray}
&& \hspace{0.0cm}\BR_{\jpsi\to \pip\pim\pip\pim}\nonumber \\
&=& (3.67\pm 0.16_{stat}\pm 0.08_{syst}\pm 0.09_{ext})\times 10^{-3}.
\end{eqnarray}
The external uncertainty (denoted {\em ext}) is dominated by the uncertainty of $\Gamma_{ee}^{\jpsi}$~\cite{PDG10}. This measurement agrees with the current PDG~\cite{PDG10} value $(3.55\pm0.23)\times 10^{-3}$.

A clear $\psitwos\to\pip\pim\mumu$ peak in the data of Fig.~\ref{psi2s}(a) is visible. Because the selection efficiency for $\pip\pim\mumu$ is the same as for $\pip\pim\pipi$, we can extract the branching fraction $\BR_{\psitwos\to\jpsi\pip\pim}$ with $\jpsi\to\mup\mun$ from a simple fit to the $\pipi\pipi$ mass distribution. Figure~\ref{psi2s_fit}(b) shows the invariant mass distribution under the $\pipi\pipi$ hypothesis in the \psitwos mass region.

The measured width of the \psitwos ($\approx20\mev$) is, as in the case of the \jpsi, dominated by the track momentum resolution, the intrinsic width of the \psitwos being $\Gamma_\psitwos=317\kev$~\cite{PDG10}. The effect of using the $\pi$ mass hypothesis in the $\mu$ track fit is negligible. The peak is described by the sum of two Gaussian distributions and the non-resonant $\pip\pim\pip\pim$ contribution with a linear function. The area under the peak is used to determine the $\psitwos\to \jpsi\pip\pim$ branching fraction according to the following equation:
\begin{eqnarray}
&& \hspace{0.0cm}\BR_{\psitwos\to \jpsi\pip\pim}\cdot \BR_{\jpsi\to\mup\mun}\cdot \sigma^{\psitwos}_{int} \nonumber \\ 
&=& \frac{N(\psitwos\to\pip\pim\mup\mun)}{d{\cal L}/dE\cdot\epsilon_{MC}} \nonumber \\ 
&=& (84.7\pm 2.2_{stat}\pm 1.8_{syst})\mevcc\nb
\end{eqnarray}
and
\begin{eqnarray}
&&\hspace{0.0cm}\BR_{\psitwos\to \jpsi\pip\pim}\cdot \BR_{\jpsi\to\mup\mun}\cdot \Gamma^{\psitwos}_{ee}\nonumber \\
&=& \frac{N(\psitwos\to\pip\pim\mup\mun)\cdot M^2_{\psitwos}}{6\pi^2\cdot d{\cal L}/dE\cdot\epsilon_{MC}\cdot C}\nonumber \\
&=& (49.9\pm 1.3_{stat}\pm 1.0_{syst})\ev,
\end{eqnarray}
where the latter result uses $M_{\psitwos}=3686.09\pm0.04\mevcc$~\cite{PDG10} and the conversion constant $C=3.8938\times 10^{11} \mev^2 \nb$~\cite{PDG10}.
The statistical uncertainty corresponds to the fit uncertainty on the area of the two Gaussian distributions and the systematic uncertainty of 3$\%$ covers the systematic uncertainties in the luminosity and efficiencies. Contributions from peaking background are negligible.
The $\BR_{\jpsi\to\mup\mun}=0.0593\pm0.0006$ branching fraction is known with very high precision. Using $\Gamma^{\psitwos}_{ee}=(2.38\pm0.04)\kev$~\cite{PDG10}, the $\psitwos\to \jpsi\pip\pim$ branching fraction is determined to be:

\begin{equation}
\BR_{\psitwos\to \jpsi\pip\pim }=0.354\pm0.009_{stat}\pm0.007_{syst}\pm0.007_{ext},
\end{equation}
where the external uncertainty is dominated by the uncertainty for $\Gamma_{ee}^{\psitwos}$~\cite{PDG10}.
The measurement is slightly higher than the PDG~\cite{PDG10} value $\BR_{\psitwos\to \jpsi\pip\pim}=0.336\pm0.005$, but agrees within the uncertainties. Our result is comparable in precision to the individual results used to determine the PDG average and agrees well with the most recent CLEO measurement $\BR_{\psitwos\to \jpsi\pip\pim }= 0.3504\pm0.0007\pm0.0077$~\cite{cleojpsipipimumu}.

\subsection{Scan for additional resonances}

\begin{figure}[t]
\epsfxsize=0.5\textwidth
\center{\epsfbox{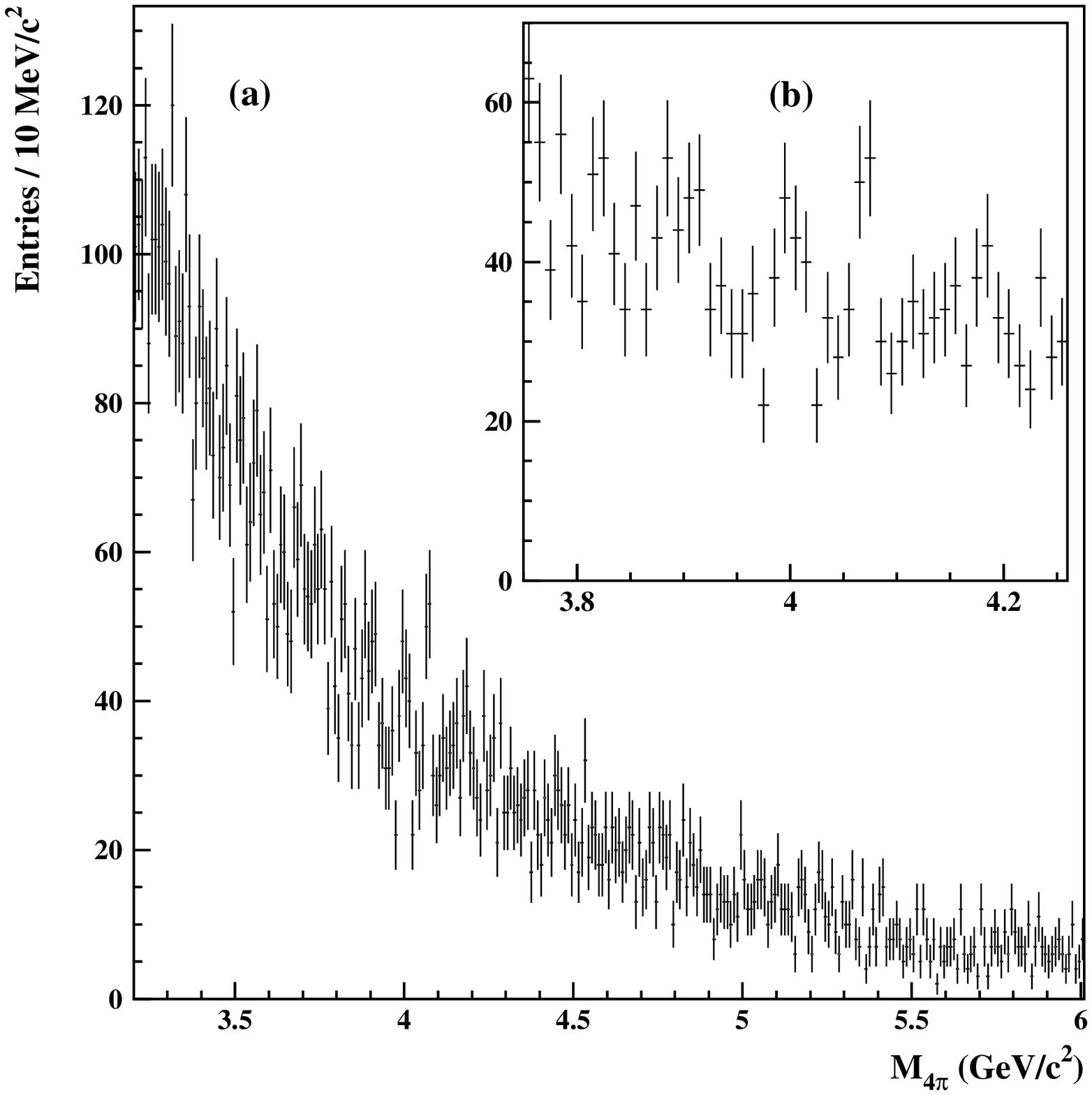}}
\caption {Invariant $\pipi\pipi$ mass distribution for the data in the invariant mass range $3.2\gevcc<M_{4\pi}<6.0\gevcc$ (a) and $3.75\gevcc<M_{4\pi}<4.25\gevcc$ (inset, b).}
\label{massscan}
\end{figure}

Fig.~\ref{massscan}(a) displays the $M_{4\pi}$ distribution for the data in the high invariant mass region. No clear signal can be identified. There is a hint of structure just above $4\gevcc$. The inset, Fig.~\ref{massscan}(b), shows this feature in more detail. 

\section{Summary}
\label{sec:Summary}

In this paper, we present a measurement of the $\pip\pim\pip\pim$ cross section at effective center-of-mass energies below 4.5\gev, using $\ep\en$ events with ISR collected in the vicinity of the \FourS resonance. We achieve overall uncertainties of 2.4\% in the peak region defined by $1.1\gevcc<M_{4\pi}<2.2\gevcc$, 10.7\% below 1.1\gevcc, 5.5\% above 2.8\gevcc and 8.5\% above 4.0\gevcc. These cross section results are much more precise than the corresponding ones based on energy scans.

The resulting contribution of the $\sigma(\ep\en\to\pipi\pipi)$ cross section to the anomalous magnetic moment of the muon $a_{\mu}^{had}(\ep\en\to\pipi\pipi)$ is evaluated according to the method described in Ref.~\cite{hvptools} in the CM energy region between 0.6 and 1.8\gev:
\begin{equation}
a_\mu^{had}(\pipi\pipi)=(13.64\pm0.03_{stat}\pm0.36_{syst})\times10^{-10}.
\end{equation}
The cross section shows evidence of resonant substructure, with preferred quasi-two-body production of $a_{1}(1260)\pi$. There is an indication of a $f_{2}(1270)\rho(770)$ contribution to the final state. A detailed understanding of the four-pion final state requires additional information from states such as $\pip\pim\piz\piz$.

The ISR events allow a study of \jpsi and \psitwos production. We measure the product of decay branching fractions and the $\ep\en$ width of the \jpsi with the best accuracy to date, with the results:

\begin{eqnarray}
&&\hspace{0.0cm}\BR_{\jpsi\to \pip\pim\pip\pim}\cdot \Gamma^{\jpsi}_{ee}\nonumber\\ 
&=& (20.4\pm0.9_{stat}\pm 0.4_{syst})\ev\\
&&\hspace{0.0cm}\BR_{\psitwos\to \jpsi\pip\pim}\cdot \BR_{\jpsi\to\mup\mun}\cdot \Gamma^{\psitwos}_{ee}\nonumber\\ 
&=& (49.9\pm 1.3_{stat}\pm1.0_{syst})\ev.\\
\nonumber
\end{eqnarray}


\section{Acknowledgments}
\label{sec:Acknowledgments}
\input acknowledgements

\input{biblio}
\end{document}

%% file: authors_jul2011.tex
%
\author{J.~P.~Lees}
\author{V.~Poireau}
\author{V.~Tisserand}
\affiliation{Laboratoire d'Annecy-le-Vieux de Physique des Particules (LAPP), Universit\'e de Savoie, CNRS/IN2P3,  F-74941 Annecy-Le-Vieux, France}
\author{J.~Garra~Tico}
\author{E.~Grauges}
\affiliation{Universitat de Barcelona, Facultat de Fisica, Departament ECM, E-08028 Barcelona, Spain }
\author{M.~Martinelli$^{ab}$}
\author{D.~A.~Milanes$^{a}$}
\author{A.~Palano$^{ab}$ }
\author{M.~Pappagallo$^{ab}$ }
\affiliation{INFN Sezione di Bari$^{a}$; Dipartimento di Fisica, Universit\`a di Bari$^{b}$, I-70126 Bari, Italy }
\author{G.~Eigen}
\author{B.~Stugu}
\affiliation{University of Bergen, Institute of Physics, N-5007 Bergen, Norway }
% ...
\collaboration{The \babar\ Collaboration}
\noaffiliation

%% file: xs_table.tex
\begin{table*}[h!]
\caption{\label{xs_sum}Summary of $\ep\en\to\pipi\pipi$ cross section measurement. {\em Dressed} (with VP) and {\em undressed} (without VP) cross sections are presented with statistical uncertainties only.}
\begin{ruledtabular}
\begin{tabular}{ c c c|c c c|c c c}
 $E_{CM} (\mev)$ & $\sigma_{4\pi}^{dressed} (\nb)$ & $\sigma_{4\pi}^{undressed} (\nb)$ & $E_{CM} (\mev)$ & $\sigma_{4\pi}^{dressed} (\nb)$ & $\sigma_{4\pi}^{undressed} (\nb)$ &$E_{CM} (\mev)$ & $\sigma_{4\pi}^{dressed} (\nb)$ & $\sigma_{4\pi}^{undressed} (\nb)$ \\
\hline
\input{xs_table2_with_gcor_11boarder.tex}

\end{tabular}
\end{ruledtabular}
\end{table*}

%% file: xs_table2_with_gcor_11boarder.tex
   612.5 & $    0.02\pm    0.01$ & $    0.02\pm    0.01$ & 1912.5 & $    7.17\pm    0.14$ & $    6.90\pm    0.13$ &  3212.5 & $    0.50\pm    0.03$ & $    0.47\pm    0.03$\\  
   637.5 & $    0.04\pm    0.02$ & $    0.04\pm    0.02$ & 1937.5 & $    6.93\pm    0.13$ & $    6.67\pm    0.13$ &  3237.5 & $    0.49\pm    0.03$ & $    0.46\pm    0.03$\\
   662.5 & $    0.02\pm    0.01$ & $    0.02\pm    0.01$ & 1962.5 & $    6.54\pm    0.13$ & $    6.30\pm    0.13$ &  3262.5 & $    0.48\pm    0.03$ & $    0.45\pm    0.03$\\
   687.5 & $    0.01\pm    0.01$ & $    0.01\pm    0.01$ & 1987.5 & $    6.04\pm    0.12$ & $    5.82\pm    0.12$ &  3287.5 & $    0.49\pm    0.03$ & $    0.47\pm    0.03$\\
   712.5 & $    0.02\pm    0.01$ & $    0.02\pm    0.01$ & 2012.5 & $    6.18\pm    0.13$ & $    5.95\pm    0.12$ &  3312.5 & $    0.47\pm    0.03$ & $    0.45\pm    0.03$\\
   737.5 & $    0.03\pm    0.01$ & $    0.03\pm    0.01$ & 2037.5 & $    5.66\pm    0.12$ & $    5.45\pm    0.12$ &  3337.5 & $    0.44\pm    0.03$ & $    0.42\pm    0.03$\\
   762.5 & $    0.05\pm    0.02$ & $    0.05\pm    0.02$ & 2062.5 & $    5.68\pm    0.12$ & $    5.47\pm    0.12$ &  3362.5 & $    0.44\pm    0.03$ & $    0.42\pm    0.03$\\
   787.5 & $    0.11\pm    0.03$ & $    0.10\pm    0.02$ & 2087.5 & $    5.34\pm    0.12$ & $    5.14\pm    0.11$ &  3387.5 & $    0.40\pm    0.03$ & $    0.38\pm    0.03$\\
   812.5 & $    0.11\pm    0.02$ & $    0.10\pm    0.02$ & 2112.5 & $    4.92\pm    0.11$ & $    4.73\pm    0.11$ &  3412.5 & $    0.38\pm    0.03$ & $    0.36\pm    0.03$\\
   837.5 & $    0.12\pm    0.03$ & $    0.12\pm    0.02$ & 2137.5 & $    4.83\pm    0.11$ & $    4.64\pm    0.11$ &  3437.5 & $    0.38\pm    0.03$ & $    0.36\pm    0.03$\\
   862.5 & $    0.17\pm    0.03$ & $    0.16\pm    0.03$ & 2162.5 & $    4.59\pm    0.11$ & $    4.41\pm    0.10$ &  3462.5 & $    0.36\pm    0.03$ & $    0.34\pm    0.02$\\
   887.5 & $    0.26\pm    0.04$ & $    0.25\pm    0.03$ & 2187.5 & $    4.28\pm    0.10$ & $    4.12\pm    0.10$ &  3487.5 & $    0.30\pm    0.02$ & $    0.28\pm    0.02$\\
   912.5 & $    0.33\pm    0.04$ & $    0.32\pm    0.04$ & 2212.5 & $    3.72\pm    0.10$ & $    3.58\pm    0.09$ &  3512.5 & $    0.35\pm    0.03$ & $    0.33\pm    0.02$\\
   937.5 & $    0.57\pm    0.05$ & $    0.55\pm    0.05$ & 2237.5 & $    3.72\pm    0.09$ & $    3.57\pm    0.09$ &  3537.5 & $    0.31\pm    0.02$ & $    0.29\pm    0.02$\\
   962.5 & $    0.71\pm    0.06$ & $    0.69\pm    0.05$ & 2262.5 & $    3.53\pm    0.09$ & $    3.39\pm    0.09$ &  3562.5 & $    0.33\pm    0.02$ & $    0.31\pm    0.02$\\
   987.5 & $    0.89\pm    0.06$ & $    0.86\pm    0.06$ & 2287.5 & $    3.26\pm    0.09$ & $    3.13\pm    0.08$ &  3587.5 & $    0.29\pm    0.02$ & $    0.28\pm    0.02$\\
  1012.5 & $    1.20\pm    0.07$ & $    1.23\pm    0.07$ & 2312.5 & $    3.18\pm    0.09$ & $    3.06\pm    0.08$ &  3612.5 & $    0.27\pm    0.02$ & $    0.26\pm    0.02$\\
  1037.5 & $    1.61\pm    0.08$ & $    1.51\pm    0.08$ & 2337.5 & $    3.06\pm    0.08$ & $    2.94\pm    0.08$ &  3637.5 & $    0.26\pm    0.02$ & $    0.25\pm    0.02$\\
  1062.5 & $    2.17\pm    0.09$ & $    2.06\pm    0.09$ & 2362.5 & $    2.97\pm    0.08$ & $    2.86\pm    0.08$ &  3662.5 & $    0.22\pm    0.02$ & $    0.22\pm    0.02$\\
  1087.5 & $    3.29\pm    0.11$ & $    3.14\pm    0.11$ & 2387.5 & $    2.59\pm    0.08$ & $    2.48\pm    0.07$ &  3687.5 & $    0.29\pm    0.02$ & $    0.13\pm    0.02$\\
  1112.5 & $    4.49\pm    0.13$ & $    4.31\pm    0.12$ & 2412.5 & $    2.47\pm    0.08$ & $    2.38\pm    0.07$ &  3712.5 & $    0.23\pm    0.02$ & $    0.21\pm    0.02$\\
  1137.5 & $    5.95\pm    0.14$ & $    5.72\pm    0.14$ & 2437.5 & $    2.30\pm    0.07$ & $    2.21\pm    0.07$ &  3737.5 & $    0.26\pm    0.02$ & $    0.24\pm    0.02$\\
  1162.5 & $    7.37\pm    0.16$ & $    7.09\pm    0.15$ & 2462.5 & $    2.25\pm    0.07$ & $    2.16\pm    0.07$ &  3762.5 & $    0.25\pm    0.02$ & $    0.23\pm    0.02$\\
  1187.5 & $    8.84\pm    0.17$ & $    8.51\pm    0.17$ & 2487.5 & $    2.11\pm    0.07$ & $    2.02\pm    0.07$ &  3787.5 & $    0.21\pm    0.02$ & $    0.20\pm    0.02$\\
  1212.5 & $   10.79\pm    0.19$ & $   10.40\pm    0.18$ & 2512.5 & $    2.03\pm    0.07$ & $    1.95\pm    0.07$ &  3812.5 & $    0.19\pm    0.02$ & $    0.18\pm    0.02$\\
  1237.5 & $   12.62\pm    0.20$ & $   12.17\pm    0.20$ & 2537.5 & $    1.87\pm    0.07$ & $    1.80\pm    0.06$ &  3837.5 & $    0.18\pm    0.02$ & $    0.17\pm    0.02$\\
  1262.5 & $   14.56\pm    0.22$ & $   14.05\pm    0.21$ & 2562.5 & $    1.71\pm    0.06$ & $    1.65\pm    0.06$ &  3862.5 & $    0.18\pm    0.02$ & $    0.17\pm    0.02$\\
  1287.5 & $   16.39\pm    0.23$ & $   15.83\pm    0.22$ & 2587.5 & $    1.85\pm    0.06$ & $    1.77\pm    0.06$ &  3887.5 & $    0.21\pm    0.02$ & $    0.20\pm    0.02$\\
  1312.5 & $   19.06\pm    0.25$ & $   18.41\pm    0.24$ & 2612.5 & $    1.79\pm    0.06$ & $    1.72\pm    0.06$ &  3912.5 & $    0.20\pm    0.02$ & $    0.19\pm    0.02$\\
  1337.5 & $   21.14\pm    0.26$ & $   20.42\pm    0.25$ & 2637.5 & $    1.62\pm    0.06$ & $    1.56\pm    0.06$ &  3937.5 & $    0.15\pm    0.02$ & $    0.14\pm    0.02$\\
  1362.5 & $   23.37\pm    0.27$ & $   22.59\pm    0.26$ & 2662.5 & $    1.43\pm    0.06$ & $    1.37\pm    0.05$ &  3962.5 & $    0.14\pm    0.02$ & $    0.14\pm    0.01$\\
  1387.5 & $   25.76\pm    0.28$ & $   24.90\pm    0.28$ & 2687.5 & $    1.31\pm    0.05$ & $    1.26\pm    0.05$ &  3987.5 & $    0.16\pm    0.02$ & $    0.16\pm    0.02$\\
  1412.5 & $   27.53\pm    0.29$ & $   26.61\pm    0.29$ & 2712.5 & $    1.30\pm    0.05$ & $    1.26\pm    0.05$ &  4012.5 & $    0.17\pm    0.02$ & $    0.16\pm    0.02$\\
  1437.5 & $   29.95\pm    0.30$ & $   28.96\pm    0.30$ & 2737.5 & $    1.21\pm    0.05$ & $    1.16\pm    0.05$ &  4037.5 & $    0.12\pm    0.01$ & $    0.11\pm    0.01$\\
  1462.5 & $   30.32\pm    0.31$ & $   29.32\pm    0.30$ & 2762.5 & $    1.17\pm    0.05$ & $    1.13\pm    0.05$ &  4062.5 & $    0.20\pm    0.02$ & $    0.19\pm    0.02$\\
  1487.5 & $   32.04\pm    0.31$ & $   30.97\pm    0.30$ & 2787.5 & $    1.17\pm    0.05$ & $    1.12\pm    0.05$ &  4087.5 & $    0.13\pm    0.01$ & $    0.12\pm    0.01$\\
  1512.5 & $   30.98\pm    0.31$ & $   29.93\pm    0.30$ & 2812.5 & $    1.09\pm    0.05$ & $    1.05\pm    0.05$ &  4112.5 & $    0.14\pm    0.02$ & $    0.13\pm    0.01$\\
  1537.5 & $   30.11\pm    0.30$ & $   29.06\pm    0.29$ & 2837.5 & $    1.07\pm    0.05$ & $    1.04\pm    0.05$ &  4137.5 & $    0.14\pm    0.02$ & $    0.13\pm    0.01$\\
  1562.5 & $   28.26\pm    0.29$ & $   27.26\pm    0.28$ & 2862.5 & $    0.96\pm    0.05$ & $    0.93\pm    0.04$ &  4162.5 & $    0.14\pm    0.01$ & $    0.13\pm    0.01$\\
  1587.5 & $   26.81\pm    0.28$ & $   25.86\pm    0.27$ & 2887.5 & $    0.89\pm    0.04$ & $    0.86\pm    0.04$ &  4187.5 & $    0.15\pm    0.02$ & $    0.14\pm    0.01$\\
  1612.5 & $   24.66\pm    0.27$ & $   23.78\pm    0.26$ & 2912.5 & $    1.08\pm    0.05$ & $    1.05\pm    0.05$ &  4212.5 & $    0.11\pm    0.01$ & $    0.10\pm    0.01$\\
  1637.5 & $   22.69\pm    0.26$ & $   21.89\pm    0.25$ & 2937.5 & $    0.88\pm    0.04$ & $    0.85\pm    0.04$ &  4237.5 & $    0.13\pm    0.01$ & $    0.12\pm    0.01$\\
  1662.5 & $   20.95\pm    0.25$ & $   20.19\pm    0.24$ & 2962.5 & $    0.77\pm    0.04$ & $    0.75\pm    0.04$ &  4262.5 & $    0.13\pm    0.01$ & $    0.12\pm    0.01$\\
  1687.5 & $   18.78\pm    0.23$ & $   18.09\pm    0.22$ & 2987.5 & $    0.82\pm    0.04$ & $    0.81\pm    0.04$ &  4287.5 & $    0.13\pm    0.01$ & $    0.12\pm    0.01$\\
  1712.5 & $   17.25\pm    0.22$ & $   16.61\pm    0.21$ & 3012.5 & $    0.75\pm    0.04$ & $    0.74\pm    0.04$ &  4312.5 & $    0.11\pm    0.01$ & $    0.11\pm    0.01$\\
  1737.5 & $   15.33\pm    0.21$ & $   14.75\pm    0.20$ & 3037.5 & $    0.71\pm    0.04$ & $    0.71\pm    0.04$ &  4337.5 & $    0.11\pm    0.01$ & $    0.11\pm    0.01$\\
  1762.5 & $   13.37\pm    0.19$ & $   12.86\pm    0.19$ & 3062.5 & $    0.62\pm    0.04$ & $    0.66\pm    0.04$ &  4362.5 & $    0.09\pm    0.01$ & $    0.09\pm    0.01$\\
  1787.5 & $   11.61\pm    0.18$ & $   11.17\pm    0.17$ & 3087.5 & $    1.93\pm    0.06$ & $    2.30\pm    0.08$ &  4387.5 & $    0.10\pm    0.01$ & $    0.10\pm    0.01$\\
  1812.5 & $   10.23\pm    0.17$ & $    9.84\pm    0.16$ & 3112.5 & $    1.30\pm    0.05$ & $    1.03\pm    0.04$ &  4412.5 & $    0.11\pm    0.01$ & $    0.10\pm    0.01$\\
  1837.5 & $    8.87\pm    0.15$ & $    8.53\pm    0.15$ & 3137.5 & $    0.62\pm    0.04$ & $    0.55\pm    0.03$ &  4437.5 & $    0.09\pm    0.01$ & $    0.09\pm    0.01$\\
  1862.5 & $    7.67\pm    0.14$ & $    7.37\pm    0.14$ & 3162.5 & $    0.59\pm    0.03$ & $    0.54\pm    0.03$ &  4462.5 & $    0.09\pm    0.01$ & $    0.08\pm    0.01$\\
  1887.5 & $    7.29\pm    0.14$ & $    7.02\pm    0.13$ & 3187.5 & $    0.51\pm    0.03$ & $    0.47\pm    0.03$ &  4487.5 & $    0.10\pm    0.01$ & $    0.09\pm    0.01$\\

%% file: acknowledgements.tex
We are grateful for the 
extraordinary contributions of our \pep2\ colleagues in
achieving the excellent luminosity and machine conditions
that have made this work possible.
The success of this project also relies critically on the 
expertise and dedication of the computing organizations that 
support \babar.
The collaborating institutions wish to thank 
SLAC for its support and the kind hospitality extended to them. 
This work is supported by the
US Department of Energy
and National Science Foundation, the
Natural Sciences and Engineering Research Council (Canada),
the Commissariat \`a l'Energie Atomique and
Institut National de Physique Nucl\'eaire et de Physique des Particules
(France), the
Bundesministerium f\"ur Bildung und Forschung and
Deutsche Forschungsgemeinschaft
(Germany), the
Istituto Nazionale di Fisica Nucleare (Italy),
the Foundation for Fundamental Research on Matter (The Netherlands),
the Research Council of Norway, the
Ministry of Education and Science of the Russian Federation, 
Ministerio de Educaci\'on y Ciencia (Spain), and the
Science and Technology Facilities Council (United Kingdom).
Individuals have received support from 
the Marie-Curie IEF program (European Union) and
the A. P. Sloan Foundation.